\documentclass[conference]{IEEEtran}
\IEEEoverridecommandlockouts

\usepackage{cite}
\usepackage{amsmath,amssymb,amsfonts}
\usepackage{algorithmic}
\usepackage{graphicx}
\usepackage{textcomp}
\usepackage{xcolor}
\usepackage{pdflscape}
\usepackage{comment}
\usepackage{acro}
\usepackage{float}
\usepackage{multirow}
\usepackage{nicematrix}
\usepackage{lipsum}
\usepackage[table]{xcolor}
\usepackage[dvipsnames]{xcolor}
\definecolor{lightyellow}{RGB}{255, 242, 204}
\definecolor{lightgreen}{RGB}{217, 234, 211}
\definecolor{lightblue}{RGB}{173, 216, 230}
\definecolor{lightred}{RGB}{244, 204, 204}

\def\BibTeX{{\rm B\kern-.05em{\sc i\kern-.025em b}\kern-.08em
    T\kern-.1667em\lower.7ex\hbox{E}\kern-.125emX}}

\DeclareAcronym{sdo}{
  short=SDOs,
  long=Standards Developing Organizations,
}
\DeclareAcronym{ai}{
  short=AI,
  long=Artificial Intelligence,
}

\DeclareAcronym{xr}{
  short=XR,
  long=Extended Reality,
}

\DeclareAcronym{vr}{
  short=VR,
  long=Virtual Reality,
}

\DeclareAcronym{3g}{
    short=3G,
    long=Third Generation Cellular Networks,
}

\DeclareAcronym{ues}{
    short=UEs,
    long=User Equipments,
}

\DeclareAcronym{ran}{
    short=RAN,
    long=Radio Access Network,
}

\DeclareAcronym{cn}{
    short=CN,
    long=Core Network,
}

\DeclareAcronym{iot}{
    short=IoT,
    long=Internet of Things,
}

\DeclareAcronym{gdp}{
    short=GDP,
    long=Gross Domestic Product,
}

\begin{document}

\title{Toward Hyper-Dimensional Connectivity in Beyond 6G: A Conceptual Framework}

\author{Ekram Hossain,~\IEEEmembership{Fellow,~IEEE,} and Angelo Vera-Rivera,~\IEEEmembership{Member,~IEEE}  

\thanks{Ekram Hossain and Angelo Vera-Rivera are with the Department of Electrical and Computer Engineering, University of Manitoba, Winnipeg, Manitoba, Canada (e-mail: ekram.hossain@umanitoba.ca and angelo.verarivera@umanitoba.ca). Corresponding author: Ekram Hossain.}}

\maketitle

\begin{abstract}
Cellular wireless networks enable mobile broadband connectivity for Internet-based applications through their radio access and core network infrastructure. While Fifth-Generation (5G) cellular systems are currently being deployed, ongoing research on cellular technologies primarily focuses on Sixth-Generation (6G) networks to set the stage for developing standards for these systems. Therefore, the time has come to articulate the visions for beyond 6G (B6G) systems.  In this article, we present a visionary framework toward hyper-dimensional connectivity in B6G that enables wireless access to hyper-immersive Internet technologies. Our contributions include a conceptual framework for B6G cellular systems with jointly integrated communication, cognition, computing, and cyber-physical capabilities as core connectivity dimensions, a set of technical definitions outlining potential use cases and system-level requirements, a mapping of prospective technology enablers, and a forward-looking research agenda for B6G systems. The conceptual discussions in this article would be helpful for identifying innovation drivers, shaping long-term technical goals, and defining research agendas for the future of mobile broadband technologies.
\end{abstract}


\section{Introduction: Entering the Era of the New Hyper-Real Internet}
\begin{figure*}
    \centering
    \includegraphics[width=0.87\linewidth]{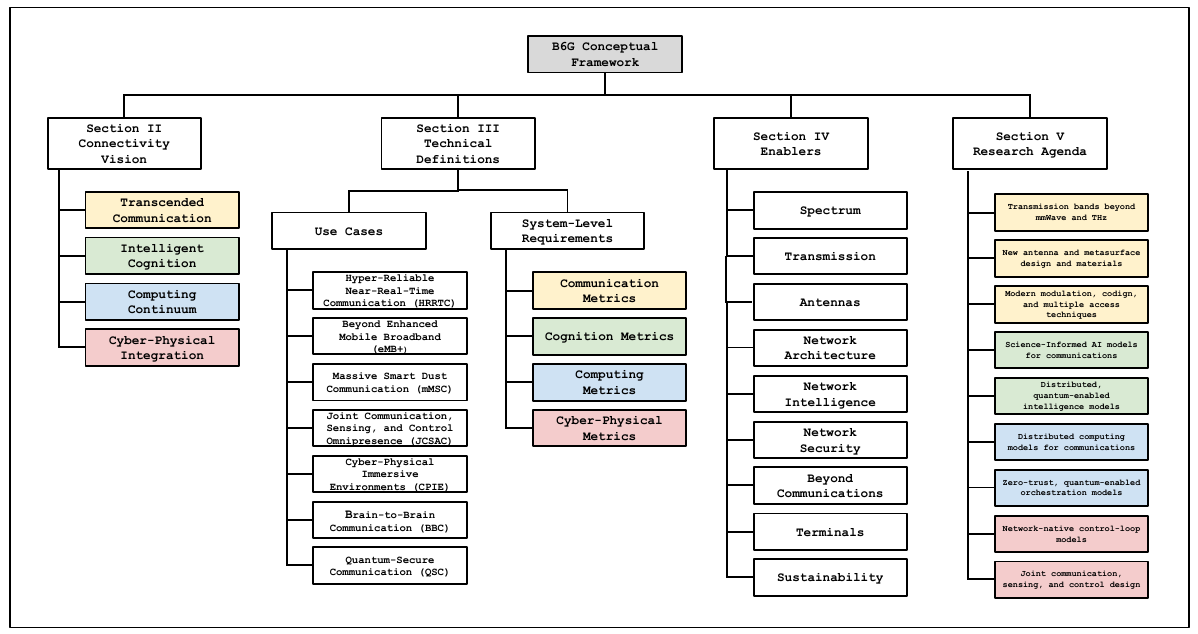}
    \caption{B6G conceptual framework outlining its core pillars: connectivity vision, technical definitions, enabling technologies, and research agenda. The figure illustrates how the four connectivity dimensions---communication (light yellow), cognition (light green), computing (light blue), and cyber-physical integration (light red)---influence other pillars of the framework.}
    \label{fig:B6ConceptualFramework}
\end{figure*}
Historically, science fiction has often anticipated technological breakthroughs. Jules Verne envisioned interplanetary travel in his 1865 novel ``From the Earth to the Moon," nearly a century before the first lunar landing in 1969. In the telecommunications domain, the 1962 cartoon ``The Jetsons" showcased a world featuring automated homes and cars decades before they emerged in the real world. More recently, Neal Stephenson introduced the Metaverse in his 1992 novel Snow Crash~\cite{Stephenson1992}, a hyper-real Internet where users immerse in a 3D virtual world through digital avatars to escape a dystopian reality.

The Internet has evolved significantly since the late 1960s, when the ARPANET project was launched to develop packet-switched computer networks~\cite{roberts1967}. Today, academia and industry leaders envision the future Internet as a platform supporting hyper-immersive experiences and multi-layered realities where users interact with digital twins in the physical world, control their data, create content, and trade assets using native cryptocurrencies in secure, decentralized environments~\cite{Zuckerberg2021, Lee_Metaverse}. In this hyper-real Internet, we are not just connecting devices, we are merging physical and digital realities into a seamless, intelligent ecosystem with multi-dimensional capabilities that could redefine how we live, work, and communicate. 

Since the mass adoption of Third-generation (3G) networks in the early 2000s, mobile broadband\footnote{As defined by the International Telecommunications Union (ITU), mobile broadband refers to mobile-cellular services that allow access to the public Internet with a download speed of at least 256 kbit/s. Mobile broadband systems include IMT-2000, IMT-Advanced, IMT-2020, and IMT-2030—colloquially known as 3G, 4G, 5G, and 6G, respectively.} have played a central role in providing wireless access to the public Internet through transmission, routing, and gateway services across their radio access and core network infrastructure. To this date, the 3rd Generation Partnership Project (3GPP)\footnote{3GPP is a consortium of telecommunications organizations and partner entities that develop specifications for cellular systems.} continues to develop specifications for Fifth-Generation (5G)-Advanced systems, marking the transition toward the standardization of the Sixth-Generation (6G) of cellular networks. 6G is planned to surpass 5G performance by enhancing transmission capacity, coverage, reliability, security, intelligence, and energy efficiency~\cite{8869705}. However, its projected capabilities may still fall short of enabling the hyper-reality Internet envisioned by academic and industry leaders.

Ongoing literature on cellular technologies primarily focuses on 6G enablers~\cite{11195786}, reflecting early-stage standardization efforts for the system. Recently, however, there has been growing interest in developing visions for beyond 6G (B6G) technologies~\cite{10963909}. Yet, most of these efforts remain communication-centric, extending 6G paradigms rather that reimagining the system. In contrast, this article presents a visionary framework toward hyper-dimensional connectivity in B6G to enable wireless access to the hyper-real Internet. While still speculative, these conceptual discussions are essential for identifying innovation drivers, shaping long-term technical goals, system performance metrics, and defining research agendas that align with the potential use cases. Our main contributions are: 
\begin{enumerate}
    \item A vision for B6G cellular networks with jointly integrated communication, cognition, computing, and cyber-physical integration as core connectivity dimensions.
    \item A set of technical definitions describing use cases and system-level requirements. 
    \item A mapping of prospective technology enablers and a forward-looking research agenda.
\end{enumerate}
The rest of the article is structured as follows (Fig.~\ref{fig:B6ConceptualFramework}). Section II presents our vision of hyper-dimensional connectivity for B6G cellular networks alongside key connectivity dimensions. Section III introduces potential use cases and requirements. Section IV examines the enabling technologies across scientific domains required to realize B6G systems. Section V outlines an forward-looking agenda with future research directions for B6G. The article concludes with final remarks and reflections. 

\section{The Vision: Hyper-Dimensional Connectivity for Cellular Networks}

Through the years, cellular wireless networks have enabled mobile broadband connectivity for a wide range of Internet-based services and business models. Today, they are foundational to the emerging Web 3.0 Internet~\cite{10736355}, unlocking sophisticated applications across numerous verticals\footnote{In economics and market theory, verticals refer to specific economic or industry sectors representing target market segments.}. These verticals include autonomous transportation, industrial IoT, online gaming, and smart city infrastructure. History shows that societal, business, and industrial demands usually drive technological evolution. It is then natural to predict that Internet technologies will follow the same trajectory. Calls for faster, more interactive, and more massive connectivity are fueled by global commerce, immersive entertainment, and the knowledge-based economy\footnote{In classical economic theory, a knowledge-based economy is an economic system in which creation, distribution, and use of knowledge is the primary driver of growth and productivity. Knowledge-based economies are highly dependent on information and communication systems.}. We predict these market segments will guide the Internet evolution in the near future, producing innovations that will enable near real time interaction between virtual and physical worlds within a planetary cyber-physical ecosystem\footnote{In systems engineering, cyber-physical systems are engineered arrangements built from the seamless integration of computational algorithms, networking models, and physical components~\cite{LeeSeshia2017}. Embedded computers and networks monitor and control the physical components, usually with feedback loops where physical components affect computations and vice versa.}  that resembles the hyper-real Internet envisioned in ``Snow Crash."

\subsection{Hyper-Dimensional Connectivity Vision}
Mobile broadband for the hyper-real Internet requires enhanced connectivity capabilities and stricter specifications than those defined for 6G in the IMT-2030 framework\cite{ITU2023}. We structure our connectivity vision for B6G around four core dimensions: (1) \textit{Transcended Communication}, (2) \textit{Cognitive Intelligence}, (3) \textit{Computing Continuum}, and (4) \textit{Cyber-Physical Integration}. Each of these connectivity dimensions is discussed next.
\subsubsection{Transcended Communication}
Transcended communication will surpass 6G communication limits by enabling peak data rates of several terabits per second (Tbps), user-experienced rates approaching 1 Tbps, spectrum efficiency near 300 bps per Hz, and sub-microsecond latency to support applications with extreme performance demands. Fully immersive services will depend on ultra-dense sensor deployments, requiring connection densities of about 1x$10^9$ devices per km$^2$, area traffic capacities in the Gbps level per m$^2$, and millimeter-level positioning accuracy. To ensure seamless integration of terrestrial and non-terrestrial platforms for global and space-based Internet access, mobility speeds must meet or exceed the 6G benchmark of 1000 km/h, with near-perfect reliability for mission-critical operations.
\subsubsection{Cognitive Intelligence}
Cognitive intelligence for B6G will expand the primitive notion of intelligence defined for 6G\footnote{In the IMT-2030 framework, AI is expected to be embedded across the network layers—from radio interfaces to the core—supporting self-monitoring, self-optimization, self-healing, and autonomous management~\cite{ITU2023}.} into a more advanced form of cognition\footnote{According to the dictionary, cognition is the act or process of knowing, including awareness, perception, reasoning, and judgment.}. In our vision, cognitive intelligence becomes holistic, integrating environmental perception, contextual learning, autonomous reasoning, and decision-making across physical and digital domains. Achieving this vision requires massive sensing through radio, optical, and physical devices that feed transparent, generalizable, and interpretable reasoning systems that adapt to highly dynamic network conditions. For truly autonomous decision-making, B6G cognition must be distributed and collaborative across cloud, edge, far-edge, and user equipment, making B6G a self-orchestrating system operating at ultra-high speeds with minimal human intervention. 
\subsubsection{Computing Continuum}
In B6G systems, computing will evolve from isolated cloud, edge, and far-edge architectures into a distributed, hyper-granular computing fabric. This fabric will integrate multiple network layers, including transmission infrastructure (e.g., compute-over-the-air technologies and stacked RISs functioning as wave-based neural networks), user equipment, and nano-scale devices. Resources and tasks will be seamlessly orchestrated with sub-millisecond allocation and secure execution in trusted environments. AI-native distributed schedulers will dynamically map workloads to resources according to latency, energy, and reliability constraints, supporting uninterrupted, computation-intensive applications. The computing continuum will form the backbone of B6G, serving as the processing substrate that sustains the convergence of communication, intelligence, and control services.
\subsubsection{Cyber-Physical Integration}
In our B6G vision, mobile broadband will evolve into a global network of distributed yet tightly integrated communication, sensing, computing, and actuation systems. This transformation will require the massive deployment and orchestration of sensors, actuators, and controllers, forming a metaphorical nervous system that can perceive, control, and act on the environment. Through cyber-physical integration, B6G will enable continuous monitoring of natural phenomena, human activity, and man-made infrastructure, followed by coordinated physical and digital responses, realizing a truly planetary cyber-physical ecosystem.   
\section{Technical Definitions: Potential Use Cases and System-level Requirements}
\begin{table*}[]
\centering
\caption{Proposed B6G metrics across the four core connectivity dimensions --communication, cognition, computing, and cyber-physical integration. Note 1: 6G benchmark metrics are defined in the IMT-2030 framework \cite{ITU2015}. Note 2: Certain metrics encompass multiple dimensions.}
\label{Tab:B6GMetrics}
\resizebox{\linewidth}{!}{
\begin{tabular}{|p{1.8cm}|p{3.5cm}|p{5.5cm}|p{4.2cm}|p{4.8cm}|p{3cm}|}
\hline
\textbf{Dimension} & \textbf{Metric} & \textbf{Details} & \textbf{6G} & \textbf{B6G} & \textbf{B6G Use Case} \\
\hline
\cellcolor{lightyellow} Communication & Peak Data Rate & Theoretical maximum data rate & Up to 20 Gbps & 5 to 10 Tbps (or higher)  & HRRTC, eMB+, BBC \\
\cellcolor{lightyellow} & User Experienced Data Rate & Average rate per user & $\sim$ 100 Mbps & 100 Gbps to 1 Tbps & HRRTC, eMB+, BBC\\
\cellcolor{lightyellow} & Spectrum Efficiency & Bits transmitted per Hz & $\sim$ 30 bps per Hz & Few hundreds of bps per Hz & eMB+, BBC\\
\cellcolor{lightyellow} & Area Traffic Capacity & Aggregate throughput per area & $\sim$ 10 Mbps per m$^2$ & Approaching the Gbps level per m$^2$ & mSDC, CPIE \\
\cellcolor{lightyellow} & Connection Density & Number of devices per area & Up to 1x10$^8$ devices per km$^2$ & Up to 1x10$^9$ devices per km$^2$ & mSDC, JCSAC, CPIE  \\
\cellcolor{lightyellow} & Mobility & Maximum supported user speed & Up to 500 km per hour & Up to 1000 km per hour & HRRTC  \\
\cellcolor{lightyellow} & Latency & Radio network transmission to end-to-end delay & Down to 1 ms & Sub-millisecond level &  HRRTC, BBC \\
\cellcolor{lightyellow} & Reliability & Successful packet delivery rate & 99.999\% ($\sim$1x10$^{-5}$ failure)  & 99.99999\% or higher ($\sim$1x10$^{-7}$ failure) & HRRTC, BBC \\
\hline
\cellcolor{lightgreen} Cognition & Positioning Accuracy & Accuracy of localization estimates & Between 1 and 10 cm & Millimeter level & JCSAC, CPIE \\
\cellcolor{lightgreen} & Probability of Detection & Target or event detection probability & Not defined & 99.9\% ($\sim$1x10$^{-3}$ failure) & JCSAC, CPIE \\
\cellcolor{lightgreen} & Inference Rate & AI inference cycles per second & Not defined & Approaching 1x10$^6$ inferences per second & JCSAC, CPIE \\
\cellcolor{lightgreen} & Inference Latency & Delay per AI inference & Not defined & Sub-millisecond level & JCSAC, CPIE \\
\hline
\cellcolor{lightblue} Computing & Task Executiuon Rate & Average number of completed tasks per second & Not defined & Approaching 1x10$^7$ tasks per second & JCSAC, CPIE \\
\cellcolor{lightblue}  & Task Execution Time & Average processing delay per task & Not defined & Sub-millisecond level & JCSAC, CPIE \\
\hline
\cellcolor{lightred} Cyber-Physical & Actuation Availability & Percentage of available control paths & Not defined & 99.999\% ($\sim$1x10$^{-5}$ failure) &JCSAC, CPIE \\
\cellcolor{lightred} & Sensing-Actuation Latency & Time between perception and actuation & Not defined & Sub-millisecond level & JCSAC, CPIE \\
\hline
Transversal  & Security & Quantum-resilient encryption strength & Not defined & 512-bit equivalent quantum-resistant & QSC\\
 & Sustainability & Throughput per unit of energy consumption& Up to 1 Gbps per Joule & Approaching 10 Gbps per Joule & All use cases \\
& Coverage & Spatial domain of network access & Global: space, air, ground, and sea & Global: space, air, ground, and sea & All use cases\\
  & Interoperability & Cross-system integration level & Most
legacy systems & Most
legacy systems & All use cases\\

\hline

        \hline
    \end{tabular}
}
\end{table*}   
Hyper-dimensional connectivity unlocks vast opportunities for advanced use cases that transcend the limits of previous cellular generations. These use cases serve as the foundation for establishing system-level requirements, identifying enabling technologies, shaping network architectures, and guiding future technical standards. We outline our envisioned use cases and corresponding requirements for B6G in the following subsections.
\subsection{Use Cases}

\subsubsection{Hyper-Reliable near real time Communication (HRRTC)}
This use case targets near real time latencies\footnote{As defined by the ITU, latency (over the air) refers to the contribution by the radio network to the time from when the source sends a packet of a certain size to when the destination receives it.} breaking millisecond levels and nearly perfect availability for applications requiring uninterrupted service and timely response for mission-critical operations at a planetary scale.
%
\subsubsection{Beyond Enhanced Mobile Broadband (eMB+)} 
This use case envisions data rates in the Tbps range, far exceeding those of previous cellular generations to support bandwidth-intensive applications. eMB+ will enable highly immersive experiences at a planetary scale, where digital twins are transmitted and rendered in real time with lifelike fidelity and environmental interaction. 
\subsubsection{Massive Smart Dust Communication (mSDC)}
This use case evolves from 6G’s ubiquitous connectivity paradigm to enable the large-scale deployment of micro- and nanoscale devices, operating at near-zero power while transmitting small data packets.
%
\subsubsection{Joint Communication, Sensing, and Control Omnipresence (JCSAC)}
This use case evolves from 6G's joint communication and sensing (JCAS) scenario. JCSAC envisions the convergence of communication, sensing, and control capabilities into an ubiquitous broadband fabric with pervasive perception, learning, reasoning and actuation integration.
%
\subsubsection{Cyber-Physical Immersive Environments (CPIE)}
This use case evolves from the immersive communication scenario in 6G. CPIE will transcend XR and AR to deliver multi-sensory interaction, including haptics, smell, and bio-signal feedback, allowing users to interact with digital twins of physical spaces creating experiences indistinguishable from reality. 
\subsubsection{Brain-to-Brain Communication (BBC)} 
This visionary B6G use case enables the direct exchange of neural information between humans, or between humans and machines, without relying on traditional communication interfaces. BBC could enable the encoding, transmission, and decoding of neural signals in near real time.
%
\subsubsection{Quantum-Secure Communication (QSC)} 
Another visionary use case for B6G is quantum-secure communication. In this scenario, B6G networks can distribute cryptographic keys using quantum principles such as superposition and entanglement, embedding confidentiality and integrity directly into the quantum-level physical layer.
%
\subsection{System-level Requirements}
\begin{figure*}[h]
    \centering
    \includegraphics[width=0.75\linewidth]{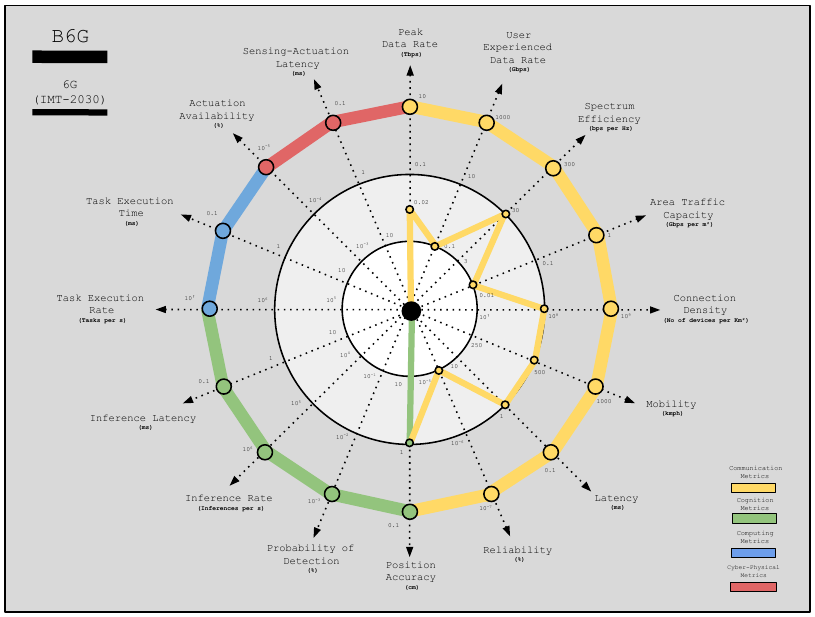}
    \caption{IMT-style illustration of B6G technical capabilities, extending the IMT-2030 (6G) communication-centric vision toward a hyper-dimensional framework that incorporates four core connectivity dimensions---communication, cognition, computing, and cyber-physical integration. B6G performance levels are represented by the thicker line, highlighting their extended range over 6G’s baseline metrics (thinner line).}
    \label{fig:B6GMetrics}
\end{figure*}
We define twenty technical requirements that capture the key capabilities of B6G networks, directly derived from the use cases discussed in the previous subsection. Some of these extend the metrics established in IMT-2020 \cite{ITU2015} and IMT-2030 \cite{ITU2023} for 5G and 6G, while others introduce new performance benchmarks unique to B6G. We expect that communication metrics such as peak data rate reach multi-terabit-per-second levels, with user-experienced rates approaching 1 Tbps. Over-the-air latency will fall to the sub-millisecond or even microsecond range, and reliability will approach seven-nines availability (99.99999\%) for performance-critical applications. Cognition requirements include sensing metrics such as millimeter-level positioning accuracy and object-detection reliability exceeding 99.9\%. To support robust cognitive services, intelligence metrics must enable inference rates near 10$^6$ inferences per second with sub-millisecond latency for learning and reasoning tasks across the network.

B6G will embed computing-as-a-service within the infrastructure, effectively functioning as a native distributed data center in the mobile environment. Task-completion rates approaching 10$^7$ computations per second with sub-millisecond processing delay represent plausible performance targets, comparable to contemporary cloud systems. For reliable cyber-physical operation, the percentage of available control paths should approach  99.999\%, with perception-to-actuation delays in the sub-millisecond range. Post-quantum security, sustainability, global coverage, and interoperability are transversal capabilities that apply across all connectivity dimensions. 

Figure 2 shows the technical capabilities of B6G in the context of the proposed hyper-dimensional connectivity framework. A summary of the proposed metrics is provided in Table \ref{Tab:B6GMetrics}, followed by an IMT-style illustration comparing the technical capabilities of 6G and B6G.
\section{Technology Enablers}
\begin{figure}
    \centering
    \includegraphics[width=\linewidth]{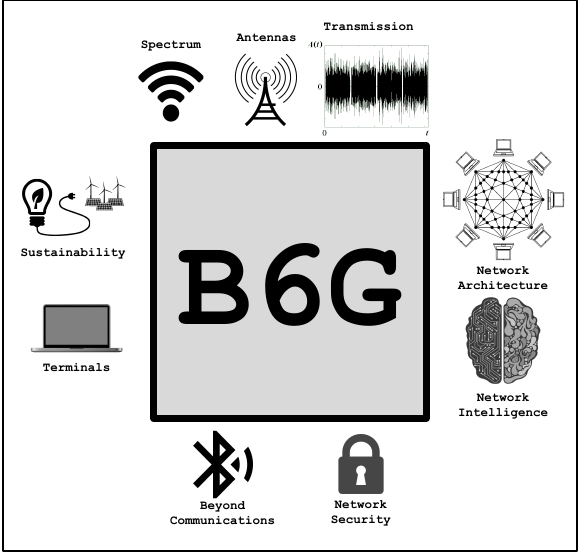}
    \caption{Technology domains for B6G highlighting key areas such as spectrum innovations, advanced antenna technologies, transmission techniques, next-generation network architectures, AI-aided communication, security and privacy frameworks, beyond-communication services, terminal device enhancements, and environmentally sustainable system design.}
    \label{fig:B6GEnablers}
\end{figure}

Realizing B6G requires the convergence of innovations across multiple scientific and engineering fields. For clarity, we group these enablers into nine core domains illustrated in Fig.~\ref{fig:B6GEnablers}. A discussion on each domain follows. 
\subsection{Spectrum}
B6G will require data transmission in frequency bands beyond mmWave and sub-terahertz (THz), extending into higher regions of the spectrum. Despite propagation challenges, these bands are essential to achieving Tbps data rates for ultra-high-speed links. Candidate regions include mmWave extensions (mmWave+), THz, optical bands (visible and infrared), and ultraviolet (UV). For certain applications---such as biological interfaces or quantum-secure communications---molecular and quantum bands offer promising alternatives for data transmission. Regardless of spectrum region, effective governance frameworks are essential for flexible licensing, intelligent allocation, dynamic aggregation, and fair spectrum sharing tailored to B6G demands.
\subsection{Antennas} 
B6G will require ultra-massive antenna arrays operating at high frequencies with {\em multi-band} and {\em multi-mode} capabilities. Integrated with intelligent active and passive metasurfaces, these antennas will provide precise signal control and enhanced spatial resolution, enabling focused energy beams for higher spectral efficiency, ultra-high data rates, and large-scale connectivity. Antennas in B6G will act not only as communication front-ends but also as sensors, energy harvesters, and control nodes, enabling advanced cyber-physical use cases. Advancing ultra-massive MIMO (umMIMO), Reconfigurable Intelligent Surfaces (RIS), and Holographic RIS will be key to achieving these goals. The umMIMO+RIS combination allows real-time manipulation of wave propagation through programmable amplitude, phase, and direction control to enhance beamforming and coverage in dense environments. To support these advancements, B6G antennas must employ innovative materials beyond conventional metals and dielectrics, offering high conductivity, mechanical flexibility, and tunable characteristics.
\subsection{Transmission}
To meet the extreme requirements for B6G, transmission schemes must evolve beyond those of 5G and 6G with more efficient multiple-access models, novel waveform design, advanced modulation, and modern coding techniques suitable for both {\em near-field} and {\em far-field} operations. Non-orthogonal multiple access (NOMA) allows multiple users to share the same resources by assigning different power levels or code sequences, significantly improving spectral efficiency. Orthogonal time–frequency space NOMA (OTFS-NOMA), in particular, is well suited for high-mobility and non-terrestrial networks due to its robustness in delay–Doppler channels. Moreover, rate-splitting multiple access (RSMA) and grant-free massive access will be key to supporting ultra-dense connectivity in B6G. For THz and optical bands, single-carrier waveforms with low peak-to-average power ratio (PAPR) will be essential to mitigate hardware nonlinearities. Techniques such as index modulation, media-based modulation, and probabilistic shaping are also gaining traction to enhance spectral and energy efficiency while reducing complexity. 
\subsection{Network Architecture}
B6G will require architectures capable of integrating highly heterogeneous network segments (terrestrial, non-terrestrial/aerial, under-water, through-the-Earth networks) on a global scale. A central element is orchestration, which manages resources, functions, and services across terrestrial and non-terrestrial infrastructures, virtualized and sliced networks, and distributed computing and control nodes spanning cloud, edge, fog, and air-wave domains. To ensure interoperability and scalability, orchestration services should be implemented as software-based modules over infrastructure-agnostic hardware. These orchestration modules should have open interfaces with standardized application programming interfaces (APIs) for vendors and third-party services. While open interfaces enhance interoperability and transparency, they also increase exposure to potential malicious actors. Achieving secure orchestration demands a shift toward zero-trust distributed frameworks~\cite{verarivera2025decentralizingtrustconsortiumblockchains}, where no entity is inherently trusted and verification is continuous at every layer to ensure that openness does not compromise resilience. Distributed trust mechanisms, supported by consortium blockchain frameworks~\cite{9210139}, will enable secure, lightweight, and transparent resource coordination. Meanwhile, quantum computing and algorithms will help solve complex optimization for consensus problems beyond classical approaches. 
\subsection{Network Intelligence}
In B6G, native artificial intelligence (AI) must evolve beyond non-transparent models toward explainable and generalizable approaches that integrate domain knowledge with hybrid data-driven and science-informed learning. Such models can interpret system dynamics, adapt to unseen scenarios, and provide transparent decision-making---crucial for trust-sensitive applications~\cite{termehchi2025koopmanbasedgeneralizationdeepreinforcement}. At the air interface, science-informed AI (ScIAI) can embed physical channel models within deep learning architectures to enable interpretable spectrum allocation, beamforming, and interference management. At the edge and fog layers, explainable lightweight distributed AI models will support human-verifiable, low-latency decisions while minimizing energy use, whereas cloud-scale AI will manage large-scale model training, orchestration, and cross-domain coordination. Large-scale AI training and inference carry a high carbon cost, therefore, AI models in B6G must adopt green intelligence principles, including model compression, efficient training, federated and transfer learning, and carbon-aware orchestration, to reduce energy-per-inference and environmental impact.
\subsection{Network Security}
Network security models rely on cryptographic schemes to protect the integrity, confidentiality, and availability of network components and data. However, classical cryptographic schemes may face vulnerabilities against emerging high-capacity quantum computers. Security in B6G must ensure quantum-resistant encryption models, integrating quantum-secure communication protocols and algorithms for robust physical-layer security. Post-quantum cryptography (PQC) also impacts privacy---a subset of security---that ensures user data is accessed with explicit consent and protect identities during data collection, transmission, processing, and storage. Quantum-proof privacy models that combine PQC with quantum key distribution (QKD) are essential for resilient mobile broadband in the quantum era~\cite{Wang2022}. 
\subsection{Beyond Communications}
B6G expands 6G's JCAS use case by fusing sensing, computation, and control into the communication fabric. B6G will not only detect spatial layouts, track motion, monitor gestures, and sense environmental conditions such as rain or pollution, but also provide real-time computation and control capabilities. By integrating distributed computing, pervasive sensor–actuator connectivity, and network-native control loops into the communication fabric, B6G will transform into a context-perceiving, action-capable nervous system for the hyper-real Internet.
\subsection{Terminals}
B6G terminals will evolve from passive endpoints into active network nodes participating in communication, sensing, computation, and control. These UEs must achieve extreme miniaturization and energy autonomy, often operating battery-free through ambient energy harvesting. They will support multi-band, multi-mode communication across mmWave, THz, optical, molecular, and quantum bands while combining real-time, high-precision sensing with control functions that translate digital commands into physical actions, enabling automated machines and cyber-physical systems that operate seamlessly within the B6G fabric.
\subsection{Sustainability}
B6G must follow green design principles across all system layers—from energy-efficient protocols and AI-driven optimization to eco-friendly materials and device architectures. Energy-aware protocols with dynamic power management will be vital to reduce consumption, while large-scale antenna, sensor, and actuator deployments should rely on energy harvesting and wireless power transfer to minimize dependence on disposable batteries. Adopting circular economy practices will further mitigate electronic waste. Beyond system design, B6G infrastructures must integrate with smart grids to coordinate energy demand, allocation, and cost optimization, as high peak loads can threaten grid stability. Ultimately, B6G must embrace carbon-aware design and governance frameworks that balance performance with environmental, economic, and societal sustainability to combat climate change.
\section{Research Agenda}
Advancing B6G will require sustained interdisciplinary research spanning physical-layer design, antennas, architectures, intelligence, control, computing, security, and sustainability. For clarity, the research agenda is organized around the four core connectivity dimensions: communication, cognition, computing, and cyber-physical integration.
\subsection{Communication}
Research in B6G communication domain must develop novel propagation models beyond THz and optical bands, higher-order modulation and adaptive coding, non-orthogonal multiple-access schemes, and efficient spectrum utilization and governance frameworks. Candidate spectrum regions include mmWave extensions (mmWave+), THz, optical, and UV bands. In antenna hardware, breakthroughs in ultra-massive MIMO, holographic RIS, and novel materials with high conductivity, mechanical flexibility, and tunable electromagnetic properties will be critical to achieving Tbps data rates and extreme connectivity densities envisioned for B6G.
\subsection{Cognition}
Research for network cognition must evolve toward distributed, transparent, explainable, and energy-efficient AI-native frameworks that enable self-learning, adaptive optimization, and decision-making. Quantum machine learning (QML) and quantum-distributed intelligence promise to accelerate large-scale optimization and enhance secure consensus mechanisms through entanglement-based coordination. Beyond task-specific intelligence, Artificial General Intelligence (AGI) in networking seeks to unify perception, reasoning, and decision-making across connectivity domains, enabling broadband systems to autonomously adapt to unseen conditions and emergent objectives.
\subsection{Computing}
Research in distributed computing for B6G must move beyond traditional cloud–edge–fog hierarchy toward a distributed, hyper-granular computing fabric that enables task processing at any point within the network. Emerging paradigms such as compute-over-the-air and stacked RISs acting as wave-based neural networks will enable in-situ analog computation within the electromagnetic domain, reducing latency and energy consumption by processing signals during transmission. From an architectural standpoint, computing models for B6G must operate within zero-trust, quantum-enabled orchestration frameworks to ensure security, scalability, and continuous verification across multi-operator, multi-domain environments. Realizing quantum-secure communication will further require advancements in quantum channel modeling, transmitters, repeaters, and memories, as well as PQC and hybrid orchestration mechanisms that enable seamless coexistence between classical and quantum networks. 
\subsection{Cyber-Physical Integration}
Research on cyber-physical integration must focus on developing joint communication, sensing, and control architectures that enable closed-loop interactions between physical and cyber systems. In B6G, these control loops will be embedded within the network fabric, allowing continuous feedback and adaptation between sensors, actuators, and intelligent agents in real time.
%
%
\section{Conclusion}
The evolution of mobile broadband has followed a clear trajectory---from communication-centric services in 4G and 5G, to the integration of sensing and primitive intelligence in 6G. 
In this article, we have presented a visionary framework that expands connectivity for B6G to include additional dimensions such as cognition, computation, and control. B6G research must progress beyond physical-layer innovations toward zero-trust integration, transparent intelligence, quantum-resilient security, and sustainable system design across the highly heterogeneous architectures. As a design philosophy, B6G networks should operate as an open, infrastructure-agnostic platform that can seamlessly orchestrate functions, services, and applications across various network types while efficiently scaling to meet dynamic resource demands. The system must be resilient to failures caused by attacks, incidents, or natural disasters, ensuring availability and self-recovery under adverse conditions. New performance metrics (beyond those for IMT-2030), which are relevant to different dimensions of hyper-connectivity, will need to be defined for multi-functional, intelligent, resilient, and secure B6G networks. 





%
\bibliographystyle{IEEEtran}
\bibliography{references}
\begingroup
\let\description\LaTeXdescription
\let\enddescription\endLaTeXdescription
\endgroup
\begin{IEEEbiographynophoto}{Ekram Hossain}(Fellow, IEEE) is a Professor and the Associate Head (Graduate Studies) of the Department of Electrical and Computer Engineering, University of Manitoba, Canada. He is a Member of the College of the Royal Society of Canada, and a Fellow of the Canadian Academy of Engineering and the Engineering Institute of Canada.
\end{IEEEbiographynophoto}
\begin{IEEEbiographynophoto}{Angelo Vera-Rivera}(Member, IEEE) is a Research Technician in the Department of Electrical and Computer Engineering, University of Manitoba, Canada. He is also an Engineer-in-Training (EIT) registered with Engineers Geoscientists Manitoba.
\end{IEEEbiographynophoto}
\end{document}